\documentclass[
  reprint,
  aps,
  prx,
  superscriptaddress,
  longbibliography
]{revtex4-2}

\usepackage{bm}
\usepackage{amsmath}
\usepackage{xcolor}
\usepackage{amssymb}
\usepackage{amsxtra,color}
\usepackage{booktabs}
\usepackage{latexsym}
\usepackage{dsfont}
\usepackage[normalem]{ulem} 
\usepackage{array}
\makeatletter
\@ifundefined{selectlanguage}
  {\newcommand{\selectlanguage}[1]{}}
  {\renewcommand{\selectlanguage}[1]{}}
\makeatother

\usepackage{graphicx}
\usepackage{mathtools}
\usepackage{booktabs}

\DeclareUnicodeCharacter{0394}{\ensuremath{\Delta}}
\DeclareUnicodeCharacter{0393}{\ensuremath{\Gamma}}

\graphicspath{ {attoimages/} }   
\usepackage{braket}
\usepackage{tikz}
\usetikzlibrary{arrows}
\usepackage{pgfplots}
\usepgfplotslibrary{groupplots}
\pgfplotsset{compat=1.3}
\usepackage{lipsum}
\usepackage{color}
\usepackage{comment}

\usepackage{natbib}
\usepackage[pagewise]{lineno}

\usepackage[T1]{fontenc}
\usepackage{lmodern}
\usepackage[utf8]{inputenc} 
\usepackage{MnSymbol}	

\usepackage[unicode]{hyperref}

\definecolor{MyDarkGreen}{rgb}{0,0.6,0}
\definecolor{MyDarkBlue}{rgb}{0,0,0.8}
\definecolor{MyDarkRed}{rgb}{0.6,0,0.3}

\hypersetup{breaklinks=true, colorlinks=true,plainpages=true, linktocpage=true, linkcolor=MyDarkBlue, citecolor=MyDarkGreen, urlcolor=MyDarkRed, pdfborder={0 0 0},%
pdfauthor={},%
pdfsubject={Research article},
pdftitle={}%
}

\begin{document}
\preprint{APS/123-QED}

\title{Probing of Core Excitons in Solid NaF with Polarization-Selective Attosecond\\Time-Resolved Four-Wave Mixing Spectroscopy}

\author{Kevin Gulu \surname{Xiong}}
 \email{gkxiong@berkeley.edu}
\affiliation{Department of Chemistry, University of California, Berkeley, California 94720, USA}

\author{Rafael \surname{Quintero-Bermudez}}
\affiliation{Department of Chemistry, University of California, Berkeley, California 94720, USA}
\affiliation{Chemical Sciences Division, Lawrence Berkeley National Laboratory, Berkeley, California 94720, USA}
\affiliation{Department of Physics, University of California, Berkeley, California 94720, USA}

\author{Vincent \surname{Eggers}}
 \altaffiliation[Present address: ]{Department of Physics and Regensburg Center for Ultrafast Nanoscopy (RUN), University of Regensburg, 93040 Regensburg, Germany}
\affiliation{Department of Chemistry, University of California, Berkeley, California 94720, USA}

\author{Hugo \surname{Laurell}}
\affiliation{Department of Chemistry, University of California, Berkeley, California 94720, USA}
\affiliation{Department of Physics, University of California, Berkeley, California 94720, USA}
\affiliation{Material Sciences Division, Lawrence Berkeley National Laboratory, Berkeley, California 94720, USA}
\affiliation{Department of Physics, Lund University, Box 118, 22100 Lund, Sweden}

\author{Melody \surname{Wu}}
\affiliation{Department of Physics, University of California, Berkeley, California 94720, USA}

\author{Stephen R. \surname{Leone}}
 \email{srl@berkeley.edu}
\affiliation{Department of Chemistry, University of California, Berkeley, California 94720, USA}
\affiliation{Chemical Sciences Division, Lawrence Berkeley National Laboratory, Berkeley, California 94720, USA}
\affiliation{Department of Physics, University of California, Berkeley, California 94720, USA}

\date{\today}

\begin{abstract}
\begin{centering}
Nonlinear four-wave mixing processes are a powerful technique to unravel ultrafast dynamics in solid-state systems. Here, we employ attosecond four-wave mixing spectroscopy with one extreme ultraviolet (XUV) pump and two independently delayed, noncollinear near-infrared (NIR) probes to resolve the ultrafast decoherence of both dipole-allowed and dipole-forbidden core excitons at the Na\textsuperscript{+} L\textsubscript{2,3} edge in sodium fluoride (NaF). The decoherence times of the core excitons are observed to be much faster than the 8 fs limit of the instrument response time, which is attributed to strong exciton–phonon coupling. Furthermore, polarization control of the NIR probes (perpendicular and parallel polarizations) reveals that the bright core excitons exhibit \emph{s}-like orbital angular momentum, while dark core excitons, reached by two-photon excitation, exhibit \emph{p}-like orbital angular momentum.
\end{centering}
 \end{abstract}

\maketitle


\section{Introduction}
Ionic insulators are an excellent platform to study fundamental physical phenomena in solid-state matter. In particular, their weak Coulombic screening, manifested in low dielectric constants, enables strong attraction between photoexcited core electron and hole pairs which leads to the formation of localized Frenkel core excitons \cite{frenkel_transformation_1931}. Early investigations of such states relied primarily on static XUV absorption, emission, and photoemission spectroscopies \cite{obrien_phonon_1993,miyano_photon-excited_1994,lapeyre_photoemission_1974,tsujibayashi_two-photon_2005,obryan_structure_1940} Although rich in spectroscopic details, these studies were limited by a number of aspects including the lack of time-domain information and accessibility to optically forbidden states or real space orbital angular momentum. Estimates based on absorption and photoemission linewidths suggested that these core excitons decay on extremely short timescales, beyond the reach of conventional ultrafast techniques.

The advent of attosecond science \cite{antoine_attosecond_1996,krausz_attosecond_2009,geneaux_transient_2019,paul_observation_2001,corkum_plasma_1993} provides researchers with spectroscopic tools fast enough to interrogate core excitons on their intrinsic timescale. In particular, the development of tabletop high harmonic generation (HHG) beamlines has made it possible to generate XUV pulses with durations in the attosecond regime. Combined with few-cycle NIR pulses, attosecond pump-probe spectroscopy has become a state-of-the-art technique for investigating few- and sub-femtosecond dynamics in both gas and condensed phases \cite{harkema_controlling_2018,volkov_attosecond_2019,goulielmakis_real-time_2010,kobayashi_direct_2019}. Beyond the exceptional temporal resolution, these techniques also feature element specificity, a capability uniquely suited for probing localized core-excited states. 

Previous studies on core excitons in ionic insulators conducted with attosecond transient absorption and reflectivity spectroscopies (ATAS and ATRS, respectively) reveal that the decoherence of core excitons indeed occurs on a few-femtosecond timescale \cite{gannan_polarization-resolved_2025,moulet_soft_2017,quintero-bermudez_deciphering_2024,geneaux_attosecond_2020,lucchini_unravelling_2021,chang_coupled_2021}. Although these studies focused mostly on dipole-allowed (bright) core excitons, they also demonstrate that dipole-forbidden (dark) core excitons play a central role in shaping the transient response through strong NIR-mediated coupling with the bright core excitons. However, dark core excitons could not be directly investigated with ATAS or ATRS, a limitation overcome by attosecond noncollinear four-wave mixing spectroscopy (FWM) in solid-state systems \cite{gaynor_solid_2021}. By focusing two delayed NIR pulses in a noncollinear geometry with respect to the XUV pulse, the phase matching condition can be exploited to resonantly generate a background-free FWM signal. The temporal delays of the two NIR pulses are independently controlled, which allows for dark states to be first populated by two-photon transitions (XUV + NIR) and later probed by the second time-delayed NIR pulse. Early attosecond FWM experiments were realized in the gas phase \cite{cao_noncollinear_2016,cao_excited-state_2018,warrick_multiple_2018,fidler_autoionization_2019,ding_time-resolved_2016} and demonstrated the ability to reliably extract the ultrafast temporal dynamics of both bright and dark states in atomic systems. Core excitons present a unique opportunity to use the power of the technique in the solid-state due to their atom-like strong absorption and localization. Gaynor \textit{et al}. \cite{gaynor_solid_2021} reported the first attosecond FWM measurements on a solid-state system, resolving bright and dark core-excitonic states of NaCl at the Na\textsuperscript{+} L\textsubscript{2,3} edges. Their study inspired other attosecond FWM works in solids not only in ionic insulators \cite{rottke_probing_2022}, but also in semimetals \cite{quintero-bermudez_attosecond_2025} and semiconductors \cite{eggers_background-free_nodate} via attosecond transient grating spectroscopy.  

However, access to the shape of the real space wave functions of core excitons has remained elusive. While theory offers valuable guidance, experimental access to their spatial structure is limited. Since core excitons are strongly bound and localized, they exhibit symmetry similar to that of atomic orbitals \cite{dolso_reconstruction_2022}. The ability to experimentally resolve this orbital character thus provides a unique perspective for studying their real-space structure. Exploiting orbital symmetry through optical polarization control is a well-established idea. For example, energy transfer cross-sections in inelastic collisions between photoexcited atoms and rare gases can be manipulated by the relative angle between excited-state orbital alignment and the relative velocity vector, with the former being controlled by the polarization of the linear excitation laser beam \cite{smith_laser_1992}. Translating this idea to ATAS, Gannan \textit{et al}. \cite{gannan_polarization-resolved_2025} were able to elucidate the orbital symmetry of core excitons in LiF by rotating the relative polarization between the pump and probe pulses. 

A FWM experimental arrangement would offer additional insight to orbital character investigations as the extra NIR pulse inherently grants FWM more degrees of freedom with which to interrogate a larger selection of dynamics. Specifically, by manipulating the polarization of pump and probe pulses, one can modify the optical selection rules beyond what can be probed by ATAS to elucidate the dynamics of dark states that can only be reached by two photons, as well as the orbital symmetry and alignment of photoexcited states in gas and condensed phases. In this work, we take the attosecond FWM methodology a step further by implementing polarization control on one of the NIR beams, allowing perpendicular and parallel polarizations, which serves as a proof of concept for experiments with advanced polarization control in attosecond nonlinear spectroscopy. We investigate the prototypical ionic insulator NaF at the Na\textsuperscript{+} L\textsubscript{2,3} edges with FWM, as it represents a simple system with strong core-excitonic XUV absorptions. Using XUV and NIR pulses with parallel linear polarizations, the ultrafast decoherence decays of both bright and dark core excitons are captured and characterized, both of which take place on a timescale much faster than the instrument response of 8 fs, pointing toward a phonon-mediated decoherence mechanism. Further, the origin of the observed FWM spectral features is assessed with the help of density functional theory (DFT) and Bethe-Salpeter equation (BSE) calculations. Finally, we present FWM results with different NIR pulse polarizations to better understand the real space structure of core excitons through the lens of orbital angular momentum. 

\begin{figure*}[htbp!]
    \centering
    \includegraphics[width=1\linewidth]{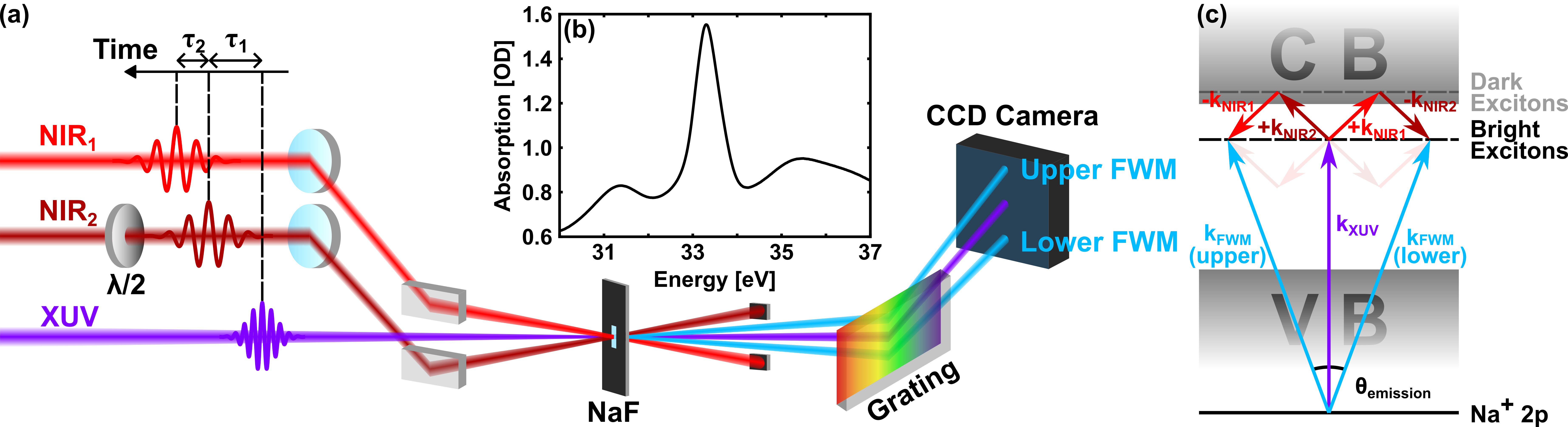}
    \caption{(a) Experimental scheme of the beamline in this work. Two NIR pulses, NIR\textsubscript{1} and NIR\textsubscript{2}, recombine noncollinearly with the XUV pulse at the NaF sample. A half waveplate ($\lambda$/2) is used to rotate the polarization of  NIR\textsubscript{2}. The XUV-NIR\textsubscript{1} and NIR\textsubscript{1}-NIR\textsubscript{2} time delays are labeled as $\tau_1$ and $\tau_2$, respectively. NIR\textsubscript{1} and NIR\textsubscript{2} are spectrally identical but are illustrated with different colors for clarity. (b) Static XUV absorption of NaF at the Na\textsuperscript{+} L\textsubscript{2,3} edge. (c) Energy level diagram and FWM coupling pathways. The vector diagram illustrates the phase-matching condition  $\bm{k}_{\mathrm{FWM}} = \bm{k}_{\mathrm{XUV}} \pm \bm{k}_{\mathrm{NIR_1}} \mp \bm{k}_{\mathrm{NIR_2}}$ of FWM. Four different pathways are possible. However, two of these are not observed due to the lack of dark core excitons at energies below the excited bright core excitons and are greyed out here. Note that the dark core excitons can be either higher or lower than the conduction band energetically depending on the exciton binding energy.}
    \label{fig:fig_1}
\end{figure*}

\section{Methods}
Details of the experimental setup can be found in Appendix\hyperref[AA]{~\ref*{AA}}. In brief, a commercial Ti:Sapphire laser is used to generate broadband NIR (500-900 nm) pulses through self-phase modulation in a hollow core fiber filled with an argon (Ar) pressure gradient. These pulses are compressed to ~5 fs using chirped mirrors and split into three arms, one of which serves as the driving field for HHG that generates attosecond XUV pulses. The two other arms, hereafter denoted as NIR\textsubscript{1} and NIR\textsubscript{2}, recombine with the XUV arm in a noncollinear geometry on a 35 nm thick polycrystalline NaF sample, which was deposited on a 10 nm thick Si\textsubscript{3}N\textsubscript{4}membrane substrate. Two piezoelectric stages are used to generate time delays between the three arms. Finally, FWM emission is spectrally dispersed by a grating and recorded by a charge-coupled device (CCD) camera.

\section{Results and Discussion}
In Fig.\hyperref[fig:fig_1]{~\ref*{fig:fig_1}(a)} the experimental scheme used to generate the FWM signal is shown. Two NIR pulses recombine noncollinearly with an XUV pulse on a 35 nm thick, polycrystalline NaF sample. The time delays of NIR\textsubscript{1} and NIR\textsubscript{2}, relative to the XUV pulse, can be independently controlled, and a half waveplate allows for the rotation of the polarization of NIR\textsubscript{2}. Within the framework of nonlinear optics, the interaction of these fields gives rise to a third-order FWM emission with a wavevector governed by the phase-matching condition,
\begin{equation}
 \bm{k}_{\mathrm{FWM}} = \bm{k}_{\mathrm{XUV}} \pm \bm{k}_{\mathrm{NIR_1}} \mp \bm{k}_{\mathrm{NIR_2}},
\end{equation}
which arises from the material’s third-order nonlinear susceptibility $\chi^{(3)}$. A vector diagram demonstrating the phase-matching geometry is shown in Fig.\hyperref[fig:fig_1]{~\ref*{fig:fig_1}(c)}. Owing to the noncollinear geometry, the FWM emission is phase-matched to propagate in two directions (labeled upper and lower wave mixing), each with a divergence angle $\theta_\mathrm{emission}$, which is given by the following under the small angle approximation,
\begin{equation}
 \theta_\mathrm{emission} = \frac{\omega_\mathrm{NIR_1}\theta_1 + \omega_\mathrm{NIR_2}\theta_2}{\omega_\mathrm{XUV}},
\end{equation}
where $\theta_1$($\theta_2$) are the angles between the propagation directions of the XUV and NIR$_1$(NIR$_2$), and $\omega_{\mathrm{XUV}}$, $\omega_{\mathrm{NIR_1}}$, and $\omega_{\mathrm{NIR_2}}$ are the angular frequencies of the corresponding pulses. This process is resonantly enhanced by NIR-mediated coupling between bright and dark core excitons, where the latter may lie either above or below the bright core excitons in energy. However, considering the bright core excitons observed in this work consist of photoexcited electrons residing in the conduction band minimum (CBM), no dark core excitons lower in energy are expected (a detailed account can be found in Section\hyperref[IIIB]{~\ref*{IIIB}}), and thus only half of the FWM pathways contribute to the measured signal in this work. 

To characterize the evolution of bright core excitons, we first populate the states with the XUV pulse and then scan over the XUV-NIR\textsubscript{1} delay ($\tau_1$) while the NIR\textsubscript{1}-NIR\textsubscript{2} delay ($\tau_2$) is fixed to zero. In this case, the decay of the FWM should therefore follow the decoherence of the bright core exciton. For this reason, we refer to these measurements hereafter as bright-state scans. Similarly, to perform a dark-state scan, $\tau_1$ is set to zero in order to initiate a two-photon XUV + NIR\textsubscript{1} transition to a dark state, followed by a scan over $\tau_2$ to extract the temporal dynamics. Importantly, the degeneracy of NIR\textsubscript{1} and NIR\textsubscript{2} in a bright-state scan dictates that upper and lower wave mixing signals in principle carry the same information. This degeneracy is broken in a dark-state scan as the interaction with NIR\textsubscript{1} must take place before that of NIR\textsubscript{2}, which is only manifested in the phase-matching of the lower wave mixing, but not the upper wave mixing. As such, the lower wave mixing is the only one that carries information for a dark-state scan and is thus what we consider in this work for both bright- and dark-state scans for consistency. 

The XUV absorption spectrum of NaF is shown in Fig.\hyperref[fig:fig_1]{~\ref*{fig:fig_1}(b)}. The peak at 33.31 eV corresponds to a strong excitonic resonance previously assigned to core excitons formed between $\Gamma$ point electrons at CBM and Na\textsuperscript{+} 2\textit{p} core hole \cite{nakai_core_1971,nakai_na_1969}. The appearance of less prominent features both within and beyond the plotted energy range have ambiguous origins and are frequently attributed to transitions at other points in the Brillouin zone \cite{nakai_core_1971,nakai_na_1969}. The constant absorption background of ~0.6 optical density (OD) is typical in XUV spectroscopy and arises from valence-to-continuum transitions.

\subsection{Ultrafast Decoherence of Core Excitons}
\begin{figure*}[htbp!]
    \centering
    \includegraphics[width=1\linewidth]{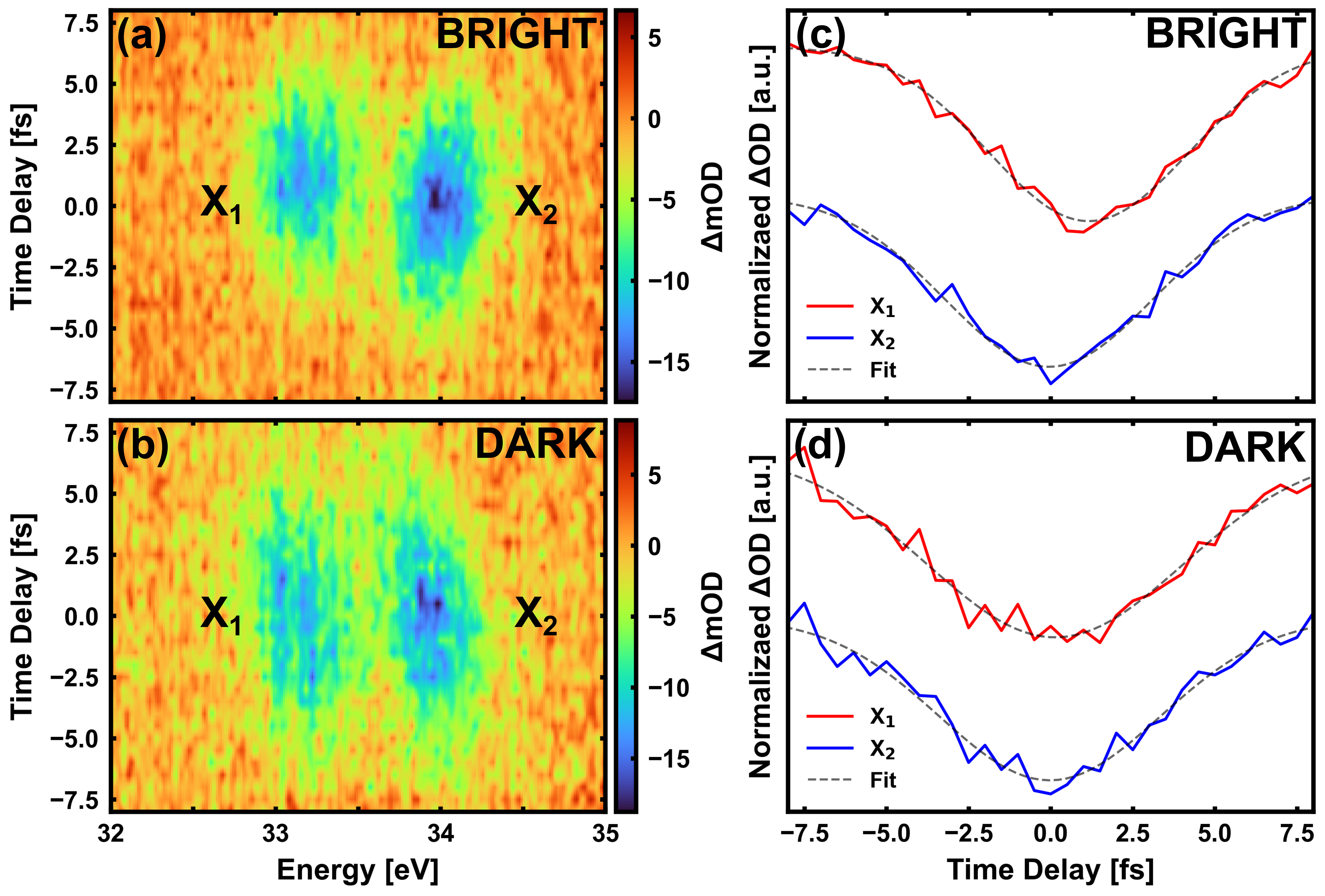}
    \caption{Evolution of core excitons observed by FWM. (a), (b) Time-resolved FWM emissions of bright and dark core excitons, respectively. Data is presented on a differential absorption scale to eliminate background. Negative time delay refers to XUV (XUV and NIR\textsubscript{1}) arriving after NIR\textsubscript{1} and NIR\textsubscript{2} (NIR\textsubscript{2}) in bright-state (dark-state) scan. Features centered at 33.18 eV in (a) and 33.15 eV in (b) are labeled X\textsubscript{1}, and features centered at 34.02 eV in (a) and 33.94 eV in (b) are labeled as X\textsubscript{2}. (c), (d) Temporal lineouts at the center energies of X\textsubscript{1} and X\textsubscript{2} for bright- and dark-state scans, respectively. Dashed traces indicate exponentially modified Gaussian fits.}
    \label{fig:fig_2}
\end{figure*}
Two distinct features (X\textsubscript{1} and X\textsubscript{2}) are observed in bright- and dark-state scans in Figs.\hyperref[fig:fig_2]{~\ref*{fig:fig_2}(a)} and\hyperref[fig:fig_2]{~\ref*{fig:fig_2}(b)} with an energy splitting of roughly 0.8 eV. These measurements were taken consecutively to ensure consistency of experimental parameters. In the bright-state scan in Fig.\hyperref[fig:fig_2]{~\ref*{fig:fig_2}(a)}, X\textsubscript{2} is launched before X\textsubscript{1} by 0.5 fs. A similar temporal trend can be observed in a separate FWM dark-state scan on NaCl \cite{gaynor_solid_2021}. Although the nature of this temporal shift is not explicitly considered or described previously, we attribute it here to temporal chirp of the NIR pulse. As X\textsubscript{1} and X\textsubscript{2} are spectrally distinct, different NIR frequency components contribute to the FWM pathways associated with each feature. A small amount of second or higher order dispersion can cause different frequency components to arrive at the sample at different times. Consequently, the effective temporal overlap required to generate each feature occurs at slightly different pump-probe delays, producing an apparent shift of 0.5 fs in the measured time zero between X\textsubscript{1} and X\textsubscript{2}. A possible doublet structure can be recognized in X\textsubscript{1} of the dark-state scan in Fig.\hyperref[fig:fig_2]{~\ref*{fig:fig_2}(b)}. While this splitting is consistent with the reported 0.16 eV spin-orbit splitting of Na\textsuperscript{+} 2\textit{p} core level \cite{wertheim_experimental_1995}, it was not definitively identified in the bright-state scan and other iterations of the same measurement. Thus, we presently refrain from assigning these features to spin-orbit components. 

Figs.\hyperref[fig:fig_2]{~\ref*{fig:fig_2}(c)} and\hyperref[fig:fig_2]{~\ref*{fig:fig_2}(d)} are lineouts of features X\textsubscript{1} and X\textsubscript{2} obtained from the bright- and dark-state scans, respectively, versus time.  Each trace was fitted using an exponentially modified Gaussian function (EMG) to account for the instrument response functions (IRFs) of the setup. The extracted full widths at half maximum (FWHM) of the Gaussian IRFs are 7.0 $\pm$ 0.7 fs and 7 $\pm$ 2 fs for X\textsubscript{1} and X\textsubscript{2}, respectively, in the bright-state scan, and 9 $\pm$ 2 fs and 8 $\pm$ 1 fs for X\textsubscript{1} and X\textsubscript{2} in the dark-state scan, respectively. Although the IRFs of the bright- and dark-state scans are similar within the experimental uncertainty, the dark-state scans exhibit longer times on average. This is not surprising, as the inclusion of an NIR pulse in the pump for dark-state scans effectively elongates the IRF. The fitted exponential decay components, on the other hand, encode core-excitonic coherence dynamics. The corresponding exponential decay constants are poorly constrained, with best-fit values always close to zero and large associated uncertainties. This suggests that the decay occurs too rapidly to be captured within the temporal resolution of the IRF, possibly in the sub-femtosecond regime. This is consistent with previous reports on other ionic insulators \cite{gannan_polarization-resolved_2025,moulet_soft_2017,quintero-bermudez_deciphering_2024,geneaux_attosecond_2020,gaynor_solid_2021}, where core excitons were found to decay in no more than a few femtoseconds.

Annihilation of the core hole through Auger-Meitner decay is a central decoherence channel of core excitons. In principle, Auger-Meitner decay sets an upper bound for the coherence lifetime as it directly reduces the core exciton population. Interestingly, most time-resolved studies of core excitons report that the decay is on a timescale much faster than what can be derived from Auger-Meitner decay \cite{gannan_polarization-resolved_2025,moulet_soft_2017,quintero-bermudez_deciphering_2024,geneaux_attosecond_2020,gaynor_solid_2021}. These findings motivate consideration of phonon-mediated decoherence as a possible explanation for this apparent discrepancy. This might seem unlikely at first as phonons generally live on the picosecond timescale while core excitons decay in femtoseconds. However, theoretical studies \cite{mahan_emission_1977,almbladh_effects_1977} indeed indicate that electron-phonon coupling initiated by the formation of the core exciton plays a significant role in modifying and broadening the linewidth observed for core-excitonic absorption and emission, and hence shortening the decoherence time. The creation of the core hole is usually accompanied by excitation of phonons that contribute to the decoherence by a factor of $e^{-\phi(t)}$, with $\phi$ being,
\begin{equation}
 \phi(t) = (M_o^2/\omega_o^2)[(2N+1)(1-\mathrm{cos}\omega_ot)-i(\omega_ot-\mathrm{sin}\omega_ot)],
\end{equation}
where $M_o$ is core hole-phonon coupling constant, $\omega_o$ is the phonon frequency, and $N$ is phonon density \cite{mahan_emission_1977}. 

The Auger lifetime of the Na\textsuperscript{+} 2\textit{p} core hole is theoretically predicted to have a linewidth of 0.153 eV \cite{green_interatomic_1987}. Moreover, it was reported experimentally with photoemission spectroscopy to be 0.10 $\pm$ 0.05 eV \cite{wertheim_experimental_1995}. The upper bound of 0.15 eV agrees with theory, and it corresponds to 7 $\pm$ 3 fs using the energy-time uncertainty relation with the linewidth $\Gamma$, $\Gamma\tau\approx\hbar$. This yields the population lifetime $T_1$, which relates to the homogeneous decoherence time $T_2$ phenomenologically as,
\begin{equation}
 \frac{1}{T_2} = \frac{1}{2T_1} + \frac{1}{T_2^*},
\end{equation}
Where $T_2^*$ is the pure dephasing caused by interaction with the environment. Assuming a perfectly isolated system ($1/T_2^* = 0$) , the decoherence time would be $2T_1 = 14$ $\pm$ 6 fs, a time much slower than the ultrafast decoherence observed in Fig.\hyperref[fig:fig_2]{~\ref*{fig:fig_2}}. Therefore, we suggest that the ultrafast decoherence of core excitons in NaF arises largely from the phonon-mediated inhomogeneous broadening, while Auger-Meitner decay plays a minor role.

\subsection{Computational Models}
\label{IIIB}
\begin{figure}[htbp!]
    \centering
    \includegraphics[width=1\linewidth]{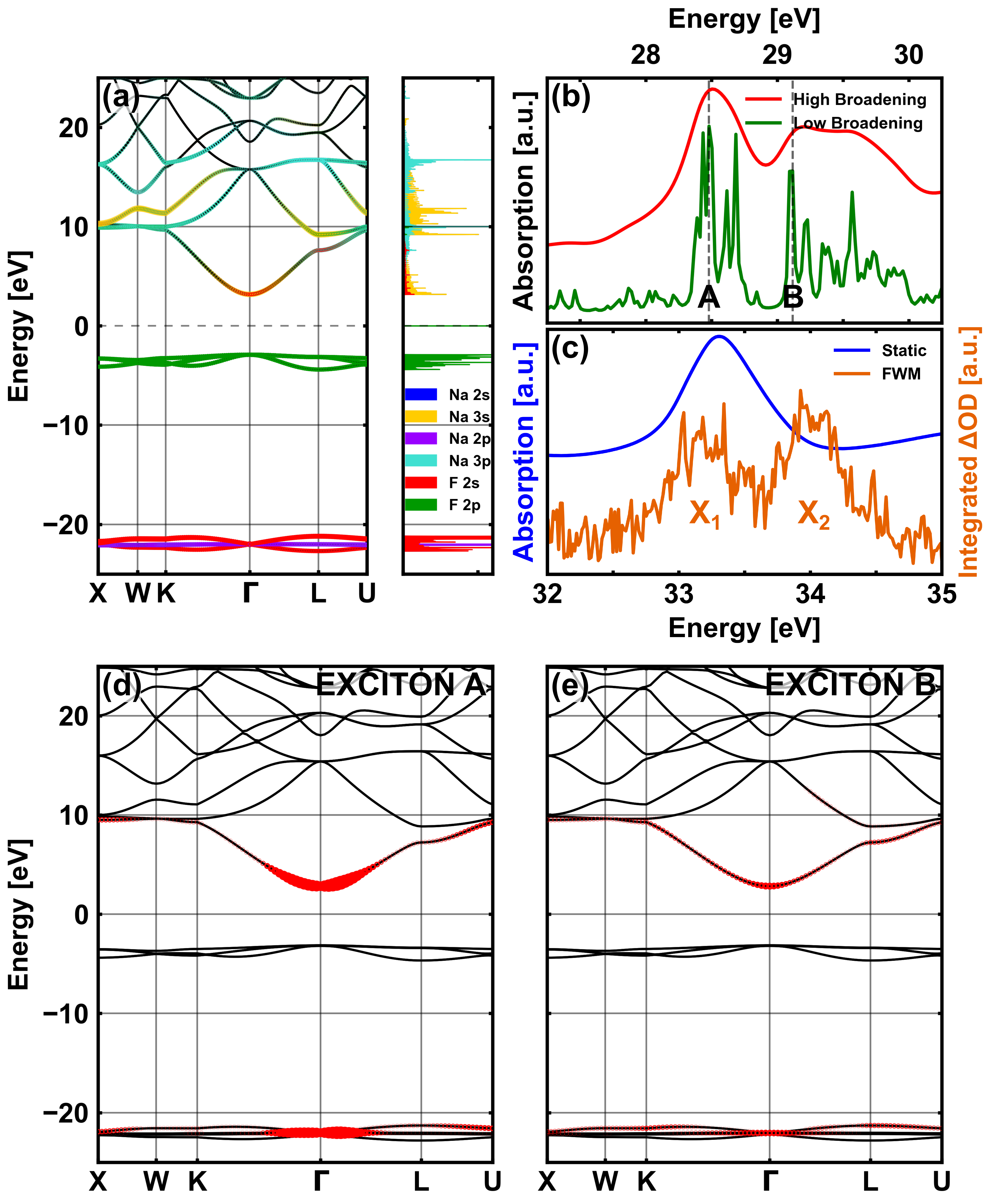}
    \caption{DFT and BSE calculations provide information on the identity of the core excitons probed by FWM. (a) PDOS of NaF calculated with the QUANTUM ESPRESSO computational package. (b) Calculated XUV absorption spectra with appropriate Gaussian broadening to match the linewidth of measured absorption (red) and negligible Gaussian broadening (green). Two characteristic excitons A and B are indicated by dashed lines. (c) Experimental XUV absorption spectrum (blue) and integrated FWM differential absorption (orange). To obtain the orange trace, Fig.\hyperref[fig:fig_2]{~\ref*{fig:fig_2}(a)} was integrated over the range of -8 fs to 8 fs. The two features are labeled X\textsubscript{1} and X\textsubscript{2} following the labeling in Fig.\hyperref[fig:fig_2]{~\ref*{fig:fig_2}}. (d), (e) Exciton weights of core excitons A and B, respectively. (b), (d), and (e) were calculated with the exciting computational package. }
    \label{fig:fig_3}
\end{figure}
To investigate the nature of the spectral features observed, we performed density functional theory calculations of the projected density of states (PDOS) in QUANTUM ESPRESSO \cite{giannozzi_quantum_2009,giannozzi_advanced_2017} (Fig.\hyperref[fig:fig_3]{~\ref*{fig:fig_3}(a)}). The CBM consists almost exclusively of Na\textsuperscript{+} 3\textit{s} character. While some F\textsuperscript{-} 2\textit{s} character can also be seen here and in previous reports \cite{gannan_polarization-resolved_2025,li_linear_1998}, it likely originates from artifacts of the linearized augmented-planewave approach used in these calculations. Since electrons are excited from the Na\textsuperscript{+} 2\textit{p} core levels, by dipole selection rules, bright core excitons are expected to have \textit{s}-like (or \textit{d}-like) orbital angular momentum and are thus formed between photoexcited Na\textsuperscript{+} 3\textit{s} electrons in CBM and Na\textsuperscript{+} 2\textit{p} core holes. It then follows that dark core excitons will have \textit{p}-like orbital angular momentum. Indeed, at higher energies in the conduction band, the onset of Na\textsuperscript{+} 3\textit{p} character becomes apparent, consistent with the expected location of dark core excitons. Note that the underestimation of the bandgap (experimentally measured to be 11.7 eV \cite{nakai_na_1969}) is a known artifact of DFT calculations and does not impact the analysis presented here.

We further employed BSE calculations to simulate the XUV absorption spectrum and exciton weights using the exciting computation package \cite{salpeter_relativistic_1951,gulans_exciting_2014}. Fig.\hyperref[fig:fig_3]{~\ref*{fig:fig_3}(b)} shows simulated XUV absorption spectra with different levels of assumed Gaussian broadening, which were added to account for phonon broadening. The red trace has a strong peak at 28.50 eV, which corresponds to the experimental excitonic resonance in Fig.\hyperref[fig:fig_3]{~\ref*{fig:fig_3}(c)} at 33.31 eV, after shifting the energy axis to correct for the underestimation of the bandgap \cite{wang_real_2019}. This peak consists of multiple excitons with significant oscillator strengths, evident from the less broadened green trace. Moreover, a broader absorption peak above 29 eV can also be identified. The exciton weights of one representative exciton from each peak (exciton A at 28.48 eV and exciton B at 29.12 eV) are illustrated in Figs.\hyperref[fig:fig_3]{~\ref*{fig:fig_3}(d)} and\hyperref[fig:fig_3]{~\ref*{fig:fig_3}(e)}. While exciton A predominantly resides at the $\Gamma$ point, exciton B is distributed more evenly over the entire Brillouin zone. This is in agreement with the assignment of previous experimental studies, where the strongest absorption peak was attributed to core excitons at the $\Gamma$ point, while the high-lying and weaker features correspond to those outside the $\Gamma$ point \cite{nakai_core_1971,nakai_na_1969}. Importantly, the spectral features of FWM cannot be mapped exactly onto the static absorption spectrum in Fig.\hyperref[fig:fig_3]{~\ref*{fig:fig_3}(c)}. Informed by these calculations, we consider several hypotheses regarding the identities of X\textsubscript{1} and X\textsubscript{2}, the two observed features in FWM. We note that the nonlinearity of the FWM process could result in the relative enhancement or reduction in signal strengths compared to linear methods. 

At first, it might be tempting to assign the two features to core hole spin-orbit components. In fact, at liquid nitrogen temperatures, it was experimentally determined that the main absorption peak centered at 33.31 eV consists of two adjacent peaks, separated by ~0.1 eV \cite{nakai_core_1971}. This doublet feature is common for sodium halides and was previously ascribed to the spin-orbit splitting \cite{nakai_core_1971,nakai_na_1969,tomita_coreexciton_2016}. As mentioned above, the spin-orbit splitting of the Na\textsuperscript{+} 2\textit{p} core level was determined to be 0.16 eV, both values being too small to explain the observed splitting of 0.8 eV. We therefore attribute feature X\textsubscript{1} to the main excitonic resonance seen in static absorption at 33.31 eV and propose that feature X\textsubscript{2} arises from another FWM-generating channel. 

As the BSE results predict a group of excitons away from the $\Gamma$ point that lie approximately 0.8 eV above the main simulated absorption peak at 28.50 eV, this aligns with an interpretation in which feature X\textsubscript{1} originates from $\Gamma$ point XUV bright core excitons while feature X\textsubscript{2} results from bright core excitons at other Brillouin-zone locations. The experimental linear absorption in Fig.\hyperref[fig:fig_3]{~\ref*{fig:fig_3}(c)} indicates that the bright core excitons that make up feature X\textsubscript{2} have significantly lower direct XUV oscillator strengths than suggested by the calculations. Nonetheless, FWM signals of X\textsubscript{1} and X\textsubscript{2} could still appear comparable in strength due to their nonlinear nature. For example, assuming X\textsubscript{1} and X\textsubscript{2} couple to the same dark core excitons, the photon energy required for X\textsubscript{2} to couple to a dark core exciton would be lower. Since the spectral intensities of the NIR pulses is higher at lower energies, this will lead to enhanced visibility in the FWM signal of X\textsubscript{2}. 

An alternative explanation for feature X\textsubscript{2} is that this signal arises due to formation of light-induced states (LISs) \cite{chen_light-induced_2012}. These states would  indicate the presence of a resonant XUV-NIR two-photon transition to a dark state, which appears only during temporal overlap. Because core excitons are extremely short-lived, an LIS can be virtually indistinguishable from a core-excitonic FWM signal since the latter is observed essentially only at temporal overlap. Prior studies on core excitons done with ATAS and ATRS frequently assigned satellite signals to LISs \cite{gannan_polarization-resolved_2025,quintero-bermudez_deciphering_2024,geneaux_attosecond_2020}, and similar reasoning can be applied in this work. The coupling scheme and phase-matching of a LIS would be similar to what is illustrated in Fig.\hyperref[fig:fig_1]{~\ref*{fig:fig_1}(c)}, except the XUV will couple the core level to an electronic virtual state instead of a bright core exciton. Depending on the bright and dark core excitons, and, by extension, the energies of XUV and NIR photons involved in this process, it could manifest as an additional feature spectrally distinct from X\textsubscript{1}. A defining signature of a LIS is that it should lie energetically at one NIR photon below or above an XUV dark state. However, the current data does not allow us to confirm this condition. 

Theory by Mahan \cite{mahan_emission_1977} suggests that the XUV emission lineshape proceeding excitation can be heavily influenced by lattice relaxation. Specifically, if the Auger lifetime is long enough to permit considerable lattice relaxation prior to emission, a Stokes shift to lower energy is expected. Interestingly, if the Auger lifetime is in an intermediate regime, theory predicts an emission spectrum with a double-humped feature, one peak with and one without a Stokes shift. This was experimentally observed in metals and insulators \cite{obrien_phonon_1993,miyano_photon-excited_1994,callcott_emission_1977,callcott_l_1978}.Given the nature of FWM as emission spectroscopy, such results exhibit striking resemblance to the measured FWM spectral features. This hypothesis, however, fails to predict the anti-Stokes energy displacement of feature X\textsubscript{2} relative to the absorption peak X\textsubscript{1}. Nevertheless, without precise knowledge of the final state accessed by the second NIR pulse, this scenario cannot be excluded. For instance, the second NIR pulse may favorably couple to feature X\textsubscript{2}, a state higher in energy compared to the one initially populated by the XUV pulse, resulting in an apparent reversal of the Stokes shift. A dedicated theoretical investigation of FWM in solids will be essential for a complete understanding.

\subsection{Polarization-Selective FWM of Core Excitons}
Assuming the bright core excitons possess \textit{s}-like symmetry, as supported by the PDOS calculation in Fig.\hyperref[fig:fig_3]{~\ref*{fig:fig_3}(a)}, the dark core excitons are expected to exhibit \textit{p}-like symmetry if they are reached with linear polarized NIR light by optical selection rules. We confirm this hypothesis experimentally and demonstrate that attosecond FWM can reveal the orbital angular momentum of core-excitonic states, and, more broadly, any electronically excited state through controlled rotation of the linear polarizations of the NIR pulses. 

\begin{figure}[htbp!]
    \centering
    \includegraphics[width=1\linewidth]{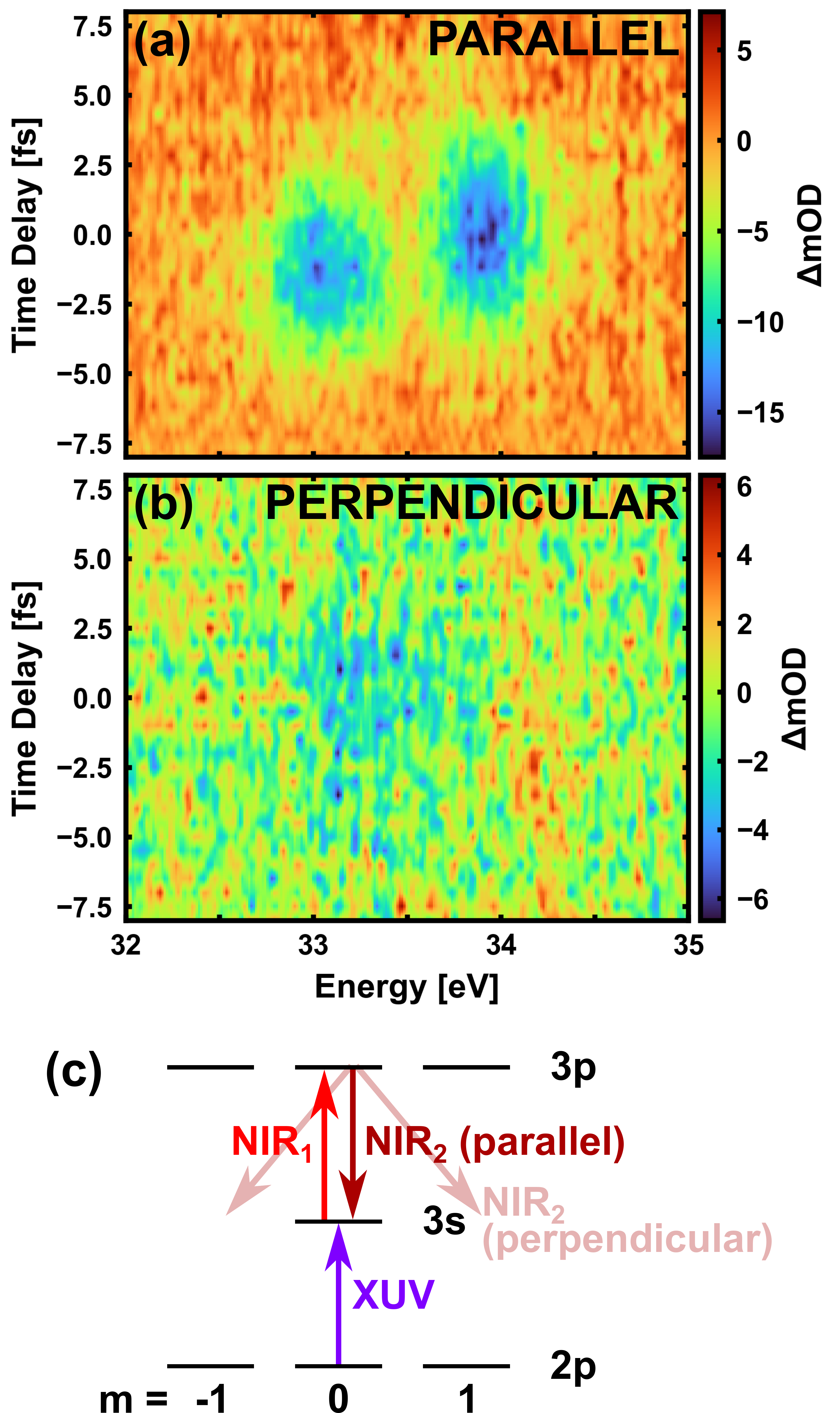}
    \caption{FWM emission measured with parallel and perpendicular NIR\textsubscript{2} polarizations. (a) Bright state FWM emission generated by XUV and NIR pulses with parallel polarizations. (b) Bright state FWM emission generated when the polarization of NIR\textsubscript{2} is perpendicular with respect to the XUV and NIR\textsubscript{1}. (c) Coupling schemes of atom-like core excitons for the parallel and the perpendicular configuration, the latter of which is forbidden.}
    \label{fig:fig_4}
\end{figure}

Under the dipole approximation, light-matter interaction occurs only when the dot product between the dipole moment and the driving electric field is nonzero: $\bm{\mu} \cdot \bm{E} \neq 0$. When the first NIR pulse excites \textit{p}-like dark core excitons, the direction of their dipole moments are inherited from the polarization of that pulse. If a second NIR pulse, polarized perpendicularly, is used to drive the transition back to \textit{s}-like bright core excitons, no light-matter interaction would take place, and no FWM emission is expected. This can also be demonstrated by considering the magnetic quantum numbers of atom-like core-excitonic orbitals in Fig.\hyperref[fig:fig_4]{~\ref*{fig:fig_4}(c)}. Further discussion below also considers possible \textit{d}-like angular momentum character, which may be present.

The perpendicular measurement in Fig.\hyperref[fig:fig_4]{~\ref*{fig:fig_4}(b)} indeed shows a disappearance of both spectral features. In comparison, a measurement was done with parallel NIR polarizations, shown in Fig.\hyperref[fig:fig_4]{~\ref*{fig:fig_4}(a)}, within the same day, where the two features are clearly visible with FWM emission signals of up to -15 $\Delta$mOD. [Although Fig.\hyperref[fig:fig_4]{~\ref*{fig:fig_4}(a)} represents a measurement identical to Fig.\hyperref[fig:fig_2]{~\ref*{fig:fig_2}(a)}, variations of experimental conditions resulted in a reversal in NIR temporal chirp, which allows X\textsubscript{1} to precede X\textsubscript{2}, opposite to the ordering observed in Fig.\hyperref[fig:fig_2]{~\ref*{fig:fig_2}(a)}.] These results confirm the \textit{s}-like and \textit{p}-like nature of the bright and dark core excitons, respectively, and represent the first experimental determination of excited state orbital symmetry using attosecond FWM. Previous work using ATAS showed similar outcomes for LiF [16].

Notably, a weak but significant FWM emission signal of -4 $\Delta$mOD can be observed in the perpendicular case centered around 33.4 eV, an energy distinct from that of the two features seen in the parallel case. As the strength of this signal is only weakly above the noise level of -2 $\Delta$mOD, it may be present in the parallel measurement, albeit difficult to isolate on the color map due to the strengths of X\textsubscript{1} and X\textsubscript{2}. This implies the mixing of orbital angular momentum of bright and dark core excitons, which breaks the respective \textit{s} and \textit{p} symmetry, allowing for the generation of FWM emission with crossed NIR polarizations. As shown by the PDOS and excitonic weight calculations in Figs.\hyperref[fig:fig_3]{~\ref*{fig:fig_3}(a)}, \hyperref[fig:fig_3]{~\ref*{fig:fig_3}(d)}, and\hyperref[fig:fig_3]{~\ref*{fig:fig_3}(e)}, the bright core excitons derive primarily from the $\Gamma$ point of the CBM, which has predominantly Na\textsuperscript{+} 3\textit{s} character. As such, the Na\textsuperscript{+} 3\textit{s} character of bright core excitons is expected to be nearly pure, whereas the dark core excitons, while predominantly \textit{p}-like, may exhibit minor mixing of orbital characteristics from \textit{s} or even \textit{d} characters to generate weak FWM emission even under the perpendicular configuration. Although no \textit{d} orbital character is obtained in the PDOS calculation in Fig.\hyperref[fig:fig_3]{~\ref*{fig:fig_3}(a)} due to the lack of consideration for 3\textit{d} orbitals in the pseudopotential used, previously published theory suggests the presence of weak Na\textsuperscript{+} \textit{d} DOS at approximately 2 eV above CBM in NaI \cite{tomita_coreexciton_2016}, despite the 8 eV gap between the Na\textsuperscript{+} 2\textit{p}3\textit{s} and Na\textsuperscript{+} 2\textit{p}3\textit{d} energy levels in the atomic counterpart \cite{wu_studies_nodate}. This is within the bandwidth of the NIR radiation used in this work, which lends validity to this interpretation.

As noted, earlier work demonstrated that the orbital angular momentum state of LiF core excitons could also be inferred by ATAS with cross-polarized pump and probe fields \cite{gannan_polarization-resolved_2025}. Here, we extend that methodology to third-order nonlinear interactions, opening up a broad range of experimental control that enables us to understand the spatiotemporal dynamics of excited states on the attosecond to femtosecond timescale. For example, rotating both NIR pulses relative to the XUV polarization could access additional selection rules and enable quantitative characterization of the extent of orbital angular momentum mixing by making measurements at varying polarization angles beyond the perpendicular configuration used in this work. Furthermore, in light of recent developments in the generation of circularly polarized XUV radiation \cite{siegrist_light-wave_2019}, techniques such as circular-dichroic ATAS \cite{drescher_attosecond_2025} became possible, which also garnered considerable theoretical interest \cite{chen_subcycle_2025}. The integration of circularly polarized light into FWM will further expand the accessible degrees of freedom and this work serves as the first step in implementing advanced polarization control in attosecond nonlinear spectroscopy. 

\section{Conclusion}
In summary, we demonstrated polarization-resolved FWM in core excitons of the ionic insulator NaF. By controlling the relative polarization of one of the NIR pulses, we directly access the orbital symmetry of bright and dark core excitons in NaF, which is determined to be \textit{s}-like and \textit{p}-like, respectively, in agreement with first-principle calculations. Furthermore, the ultrafast time resolution allowed for the investigation of the decoherence mechanism of these core excitons, which we attribute to significant exciton-phonon coupling. More broadly, this work establishes FWM as a powerful spectroscopic tool in the investigation of ultrafast dynamics in solids with unique sensitivity to orbital angular momentum character, lattice dynamics, and dipole forbidden transitions. The polarization control demonstrated in this work paves the way towards expanding the utility of XUV nonlinear and multidimensional spectroscopy for the study of complex solid-state systems.

\section*{Acknowledgments}
The authors thank Jonah Adelman, Kylie Gannan, Nicolette G. Puskar, Patrick Rupprecht, and Bokang Hou for fruitful discussions. This work was supported by the Air Force Office of Scientific Research (AFOSR) Grant Nos. FA9550-24-1-0184 and FA9550-20-1-0334. K.G.X was also partially supported by U.S. Department of Energy, Office of Science, Basic Energy Science (BES), Materials Sciences and Engineering Division under contract DE-AC02-05CH11231 within the Fundamentals of Semiconductor Nanowires Program (KCPY23). R.Q.B. acknowledges support from Natural Sciences and Engineering Research Council of Canada (NSERC) Postdoctoral Fellowship. H.L. acknowledges support from the Swedish Research Council (2023-06502) and the Sweden-America Foundation.

\section*{Data Availability}
All data and analysis codes supporting the findings of this study are publicly available at \cite{data_2026}.

\appendix
\section{Experimental Methods}
\label{AA}
\begin{figure}[htbp!]
    \centering
    \includegraphics[width=1\linewidth]{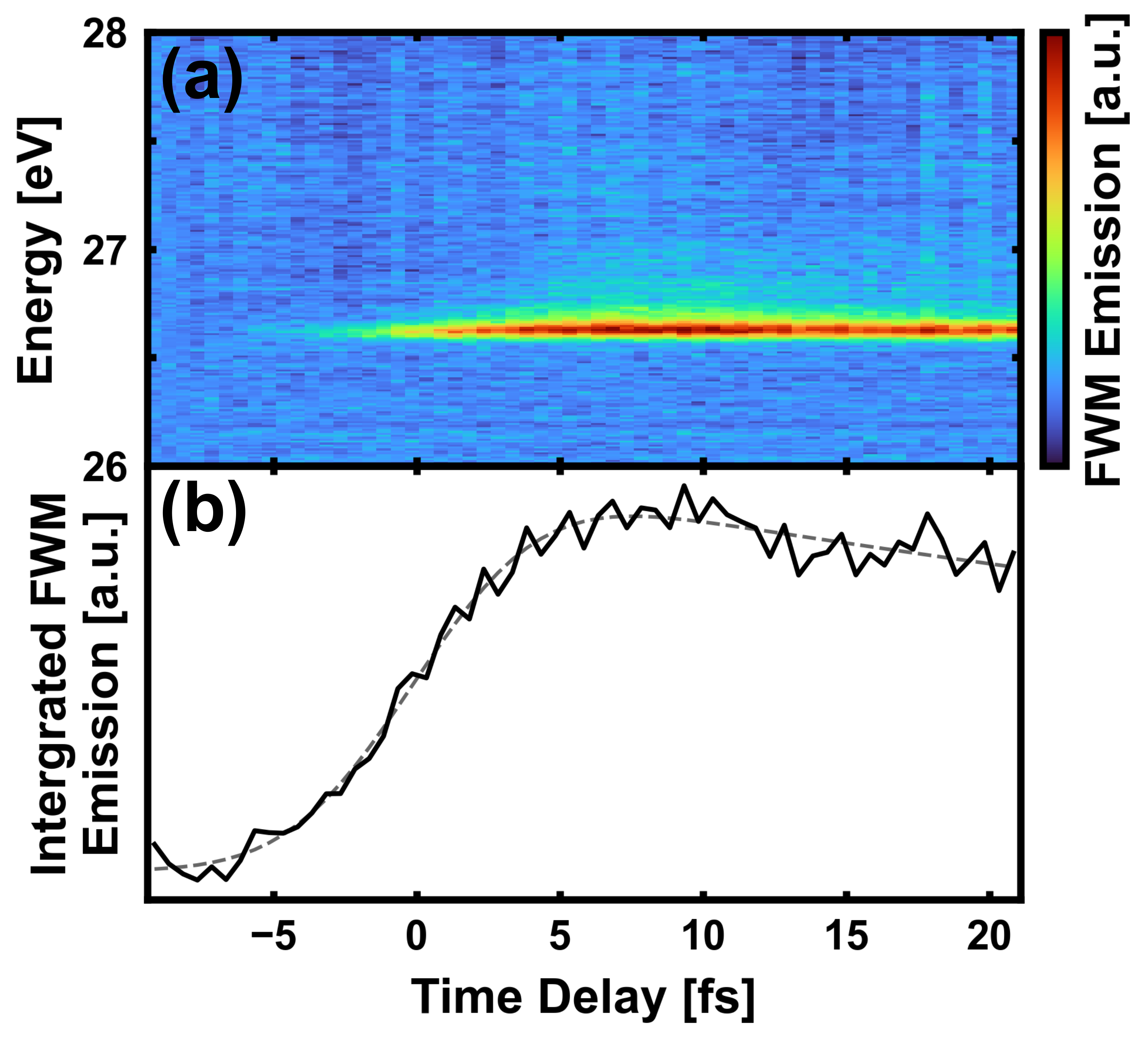}
    \caption{Example of a FWM emission scan in Ar taken immediately after NaF FWM scans. (a) FWM emission of the Ar 3\textit{s}\textsuperscript{2}3\textit{p}\textsuperscript{6} $\rightarrow$ 3\textit{s}3\textit{p}\textsuperscript{6}4\textit{p} autoionizing state. (b) Integrated FWM emission between 26.61 eV and 26.64 eV (black) and EMG fit (gray dashed). The IRF in this instance was fitted to be 8.3 $\pm$ 0.6 fs.}
    \label{fig:fig_5}
\end{figure}

The attosecond FWM beamline is driven by a Ti:sapphire laser system (Coherent Astrella USP, 800 nm center wavelength, 7 mJ maximum pulse energy, 35 fs pulse duration, 1 kHz repetition rate). Approximately 3 mJ of the laser’s output is focused with a 2.5 m focal length mirror into an Ar-gradient-filled (approximately 100 mTorr at entrance and 5 Torr at exit) hollow core fiber with 700 $\mu$m core diameter. This generates a supercontinuum spanning 500 to 900 nm through self-phase modulation. Dispersion compensation is achieved by eight pairs of broadband chirped mirrors (Ultrafast Innovations, PC1332), followed by adjustable fused silica and potassium dihydrogen phosphate wedge pairs for fine control. The resulting pulse is compressed to ~5 fs and split into two arms via a 90:10 beamsplitter. The 90\% arm is focused with a 45 cm focal length mirror into a gas cell with 25 Torr krypton with a 200 $\mu$m diameter aperture in a vacuum chamber maintained below 10\textsuperscript{-6} Torr, which undergoes HHG to produce broadband XUV pulses. A 70 nm thick Al-foil (Lebow Company) removes copropagating NIR before the XUV beam is refocused onto the sample using a gold-coated toroidal mirror. The 10\% arm is further divided into two by a 50:50 broadband beamsplitter (denoted NIR\textsubscript{1} and NIR\textsubscript{2}). A retroreflector mounted on a piezoelectric translation stage (Physik Instrumente, P-620.1CD), positioned prior to the split, controls the relative delay between the XUV pulse and both NIR pulses. An additional delay line placed in the beam path of NIR\textsubscript{2} after the split controls the relative delay between NIR\textsubscript{1} and NIR\textsubscript{2}. All FWM scans presented in this work were taken with time steps of 0.5 fs, followed by interpolation of the time delay axis in post-processing to enhance visual clarity. A zero-order half waveplate (Bernhard Halle Nachfl.) is used in NIR\textsubscript{2} for polarization control. NIR beams are focused by a 1 m focal length mirror into the vacuum chamber and onto the sample at angles of incidence of approximately $\pm$30 mrad. The pulse energy of both NIR beams are adjusted to 4.5 $\mu$J with an iris, which corresponds to an intensity of 4.0 × 10\textsuperscript{12} W/cm\textsuperscript{2}, considering a beam diameter of 240 $\mu$m at $1/e^2$. The transmitted and emitted XUV passes through another 70 nm Al-foil to filter out residual NIR. The XUV spectrum is then dispersed by a gold-coated flat-field grating (01-0639, Hitachi) and collected by a CCD camera (Princeton Instruments, Pixis XO 400-B), with a spectral resolution of 14.5 meV around the Na\textsuperscript{+} L\textsubscript{2,3} edges at 33 eV. The temporal overlap is determined by consecutively measuring the FWM of Ar 3\textit{s}3\textit{p}\textsuperscript{6}\textit{np} bright states (see Fig.\hyperref[fig:fig_5]{~\ref*{fig:fig_5}}) and fitting the response to an EMG,
\begin{equation}
 \mathrm{EMG}(t) = \frac{1}{2K} \mathrm{exp} \left( \frac{1}{2K^2} - \frac{t}{K} \right) \mathrm{erf} \left( -\frac{t-\frac{1}{K}}{\sqrt{2}} \right),
\end{equation}

\noindent where $K = \tau / \sigma$, with $\tau$ being the exponential decay constant and $\sigma$ the standard deviation of the Gaussian function. This calibration is enabled by a two-axis translation stage that allows the NaF sample to be exchanged with an Ar gas cell in the plane perpendicular to the XUV propagation direction. The known energies of these Ar resonances, reported by Madden \textit{et al}. \cite{madden_resonances_1969}, are also used to calibrate the energy axis of the spectrometer. The samples used are 35 nm polycrystalline NaF (Lebow Company) thin films deposited onto 10 nm thick Si\textsubscript{3}N\textsubscript{4}membranes (Norcada NX5050X). 

\section{Computational Methods}
The band structure and PDOS in Fig. 3(a) are calculated in QUANTUM ESPRESSO. We use the Perdew-Burke-Ernzerhof (PBE) generalized gradient approximation for exchange-correlation functional, treat ionic cores with projector augmented wave (PAW) pseudopotentials, and use a k grid of 8 × 8 × 8 for reciprocal space sampling. Excited states calculations are done with the exciting computational package. First, a ground state calculation is performed using DFT with PBE functional and a k grid of 8 × 8 × 8, where spin-orbit coupling is also included. Then, the GW approximation is applied to improve accuracy of the calculated band structure. Using the ground state results as a basis, excited state properties including exciton weights and absorption spectrum are calculated using BSE, which considers the Na 2\textit{p} and F 2\textit{s} core levels and the first six conduction band levels. 

\bibliographystyle{apsrev4-2-titles}
\bibliography{NaF_References}

\clearpage

\end{document}